\documentclass[preprint]{elsarticle}

\usepackage{lineno,hyperref}

\usepackage{xcolor}
\usepackage{soul}
\modulolinenumbers[5]
\lccode`\4`\4
\makeatletter
\def\mdseries@tt{m}
\makeatother

\usepackage{graphicx}
\usepackage{float}
\usepackage{subcaption}
\usepackage{booktabs}
\usepackage{siunitx}
\usepackage{listings}
\usepackage{multirow}
\usepackage{amsmath}
\usepackage{xcolor}

\usepackage{hyperref}

\definecolor{my_white}{rgb}{1, 1, 1}                
\definecolor{my_black}{rgb}{0, 0, 0}                
\definecolor{my_blue}{rgb}{0, 0.125, 0.376}         
\definecolor{my_green}{rgb}{0.12, 0.3, 0.17}        
\definecolor{my_violet}{rgb}{0.44, 0.16, 0.39}      

\lstdefinestyle{Bash}{
  backgroundcolor=\color{my_white},   
  basicstyle=\footnotesize,        
  breakatwhitespace=false,         
  breaklines=true,                 
  captionpos=b,                    
  commentstyle=\color{my_green},   
  frame=single,	                   
  keepspaces=true,                 
  keywordstyle=\color{my_blue},    
  language=bash,                   
  numbers=none,                    
  numbersep=5pt,                   
  numberstyle=\tiny\color{my_black},   
  rulecolor=\color{my_black},      
  showspaces=false,                
  showstringspaces=false,          
  showtabs=false,                  
  stepnumber=1,                    
  stringstyle=\color{my_violet},   
  tabsize=2,	                   
  morekeywords={git, sudo, apt, get, pip2, okular, wxparaver},             
  title=\lstname                   
}

\lstdefinestyle{PyCOMPSs}{
  backgroundcolor=\color{my_white},   
  basicstyle=\footnotesize,        
  breakatwhitespace=false,         
  breaklines=true,                 
  captionpos=None,                    
  commentstyle=\color{my_green},   
  frame=single,	                   
  keepspaces=true,                 
  keywordstyle=\color{my_blue},    
  language=Python,                 
  numbers=none,                    
  numbersep=5pt,                   
  numberstyle=\tiny\color{my_black},   
  rulecolor=\color{my_black},      
  showspaces=false,                
  showstringspaces=false,          
  showtabs=false,                  
  stepnumber=1,                    
  stringstyle=\color{my_violet},   
  tabsize=2,	                   
  morekeywords={@task, @constraint, @parallel, compss_wait_on},             
}

\title{Enabling Dynamic and Intelligent Workflows for HPC, Data Analytics, and AI Convergence}

\author[bsc]{Jorge Ejarque}
\author[bsc]{Rosa M. Badia}
\author[atos]{Loïc Albertin}
\author[cmcc]{Giovanni Aloisio}
\author[ingv]{Enrico Baglione}
\author[bsc,upc]{Yolanda Becerra}
\author[siemens]{Stefan Boschert}
\author[bsc]{Julian R. Berlin}
\author[cmcc]{Alessandro D'Anca}
\author[cmcc]{Donatello Elia}
\author[atos]{François Exertier}
\author[unitn]{Sandro Fiore}
\author[upv]{José Flich}
\author[csic,bsc]{Arnau Folch}
\author[ngi]{Steven J. Gibbons}
\author[awi]{Nikolay Koldunov}
\author[bsc]{Francesc Lordan}
\author[ingv]{Stefano Lorito}
\author[ngi]{Finn L{\o}vholt}
\author[uma]{Jorge Mac\'ias}
\author[unical]{Fabrizio Marozzo}
\author[ingv]{Alberto Michelini}
\author[bsc]{Marisol Monterrubio-Velasco}
\author[eth]{Marta Pienkowska}
\author[bsc]{Josep de la Puente}
\author[upc,bsc]{Anna Queralt}
\author[upv]{Enrique S. Quintana-Ortí}
\author[bsc]{Juan E. Rodríguez}
\author[ingv]{Fabrizio Romano}
\author[cimne,upc]{Riccardo Rossi}
\author[jsc]{Jedrzej Rybicki}
\author[psnc]{Miroslaw Kupczyk}
\author[ingv]{Jacopo Selva}
\author[unical]{Domenico Talia}
\author[ingv]{Roberto Tonini}
\author[unical]{Paolo Trunfio}
\author[ingv]{Manuela Volpe}

\address[bsc]{Barcelona Supercomputing Center (BSC)}
\address[cimne]{Centre Internacional de Mètodes Numèrics a l'Enginyeria (CIMNE)}
\address[upc]{Universitat Politècnica de Catalunya (UPC)}
\address[jsc]{J\"ulich Supercomputing Centre (JSC)}
\address[upv]{Universitat Politècnica de València (UPV)}
\address[atos]{Atos BDS R\&D HPC \& AI Software}
\address[unical]{DtoK Lab Srl}
\address[cmcc]{Centro Euro-Mediterraneo sui Cambiamenti Climatici (CMCC)}
\address[unitn]{Department of Information Engineering and Computer Science, University of Trento}
\address[uma]{Universidad de M\'alaga (UMA)}
\address[ingv]{Istituto Nazionale di Geofisica e Vulcanologia (INGV)}
\address[awi]{Alfred-Wegener-Institut Helmholtz-Zentrum f\"ur Polar- und Meeresforschung}
\address[csic]{Consejo Superior Investigaciones Cientificas (CSIC)}
\address[eth]{Eidgen\"ossische Technische Hochschule (ETH) Z\"urich}
\address[siemens]{Siemens AG}
\address[ngi]{Norwegian Geotechnical Institute (NGI)}
\address[psnc]{Poznan Supercomputing and Networking Center (PSNC)}

\begin{document}
    
    \begin{abstract}
The evolution of High-Performance Computing (HPC) platforms enables the design and execution of progressively larger and more complex workflow applications in these systems. The complexity comes not only from the number of elements that compose the workflows but also from the type of computations they perform. While traditional HPC workflows target simulations and modelling of physical phenomena, current needs require in addition data analytics (DA) and artificial intelligence (AI) tasks. However, the development of these workflows is hampered by the lack of proper programming models and environments that support the integration of HPC, DA, and AI, 
as well as the lack of tools to easily deploy and execute the workflows in HPC systems. To progress in this direction, this paper presents use cases where  complex workflows are required and investigates the main issues to be addressed for the HPC/DA/AI convergence. Based on this study, the paper identifies the challenges of a new workflow platform to manage complex workflows. Finally, it proposes a development approach for such a workflow platform addressing these challenges in two directions: first, by defining a software stack that provides the functionalities to manage these complex workflows; and second, by proposing the HPC Workflow as a Service (HPCWaaS) paradigm, which leverages the software stack to facilitate the reusability of complex workflows in federated HPC infrastructures. Proposals presented in this work are subject to study and development as part of the EuroHPC eFlows4HPC project.  
\end{abstract}


    \maketitle

\section{Introduction} 
\label{sec:introduction}

High Performance Computing (HPC) plays an increasingly important role across all scientific fields, and simulation has established itself as the third pillar, alongside theory and experiments, fostering scientific and engineering advances. This has been recognised globally leading to ambitious investments in HPC in the US, China, and Japan. Europe is not different and has created the EuroHPC Joint Undertaking, which is pooling EU and national resources in the order of billions of Euros to develop European competitive science and technology, deploying top-of-the-range exascale supercomputers. 

In addition, the recent wide availability of Big Data sources has catalyzed a data-centric science based on the intelligent analysis of these data collections and on learning techniques for gleaning the rules hidden in them. 
Such data collections may be the result from large HPC simulations, raw data from field/laboratory experiments, measurements of physical phenomena, 
gathered from the Web, and in general produced in different scientific and engineering fields.


In this sense, the scientific process has been described as consisting of three inference steps:
abduction (i.e., guessing at an explanation), deduction (i.e., determining the necessary consequences of a set of propositions), and induction (i.e., making a sampling-based generalisation). These key logical elements have been presented in~\cite{asch2018big} by the Big Data and Extreme-Scale Computing (BDEC)~\cite{web:bdec}, an international initiative that focuses on the convergence of data analytics (DA) and High-Performance Computing (HPC). While the abduction and induction involve the use of analysis and analytics processes (DA techniques), the deduction is typically an HPC process. However, the three different steps of the scientific process have been realised until now with separated methodologies and tools, with a lack of integration and common view of the whole process. The main BDEC recommendation is to address the basic problem of the separation between the two paradigms: the HPC and DA software ecosystems.
In addition, current international roadmaps, including that of BDEC, advocate for combining HPC with artificial intelligence (AI), itself tightly linked to the DA revolution. 
An additional observation is that the usage of HPC resources by scientific workflows is often conducted in a brute force manner, by submitting a large number of simulations or modelling jobs, generating a large amount of data which are then to be analysed/processed in a decoupled process. There is thus a need for smarter workflow approaches, able to leverage HPC in a more energy-efficient way but also able to carry out the different HPC, DA and AI steps in a more integrated form. 
The situation is similar in the context of industrial applications: for example, in the area of manufacturing, current technologies based on Full Order Models (FOM), developed for increasingly complex designs, generate a large amount of data that is processed in later steps to obtain Reduced Order Models (ROM) that can be used in the construction of digital twins. A more integrated approach will streamline the solution of FOM problems, paving the road toward adaptive algorithms. This, in turn, will allow faster and more reliable ROM, reducing the required simulation time and thus having a positive impact in the industry.

However, creating these new integrated workflows is not an easy task. Every HPC, DA or AI step of these workflows is often implemented as a stand-alone framework designed for a specific purpose. Developers have to dedicate a considerable effort to manage the integration of different frameworks in distinct phases of the workflow lifecycle. In the development phase, developers have to program the integration of the different workflow components implemented in a variety of programming models. In the deployment phase, different tools and frameworks must be deployed in the infrastructure. Finally, in the execution phase, the execution of all the distinct components must be orchestrated dynamically and intelligently. 
While in our approach the task graph of the workflows is generated at runtime, this is not the only source of dynamism.  For example, we also consider that the workflow may have to adapt according to specific data inputs or intermediate results; and the workflow may also need to react to failures or exceptions, cancelling parts of the workflow and/or spawning new computations on it. By intelligent, we do not only mean that the workflows include AI elements in their application components, but also that the runtime can make intelligent decisions to improve the workflow execution. These may range from automating processes to reduce human intervention to task-scheduling policies that take into account data locality, or that implement resource elasticity to improve energy efficiency. 
For these reasons, new workflow platforms enabling the design of complex applications that integrate HPC, DA, and AI processes are necessary.

These platforms should exploit the use of the HPC resources in an easy, efficient, and responsible way as well as enable the accessibility and reusability of applications to reduce the time to solution.
To this end, this paper analyses the context of HPC, DA and AI convergence and presents use cases where these complex workflows are required. Based on this analysis, the paper exposes the challenges of delivering a new workflow platform to manage complex workflows. Finally, it proposes a development approach for such workflow platform which addresses these challenges. This platform consist of two parts: a software stack that provides the functionalities to manage these complex workflows, and the HPC Workflow as a Service (HPCWaaS) concept, which leverages the software stack to facilitate the reusability of complex workflows in federated HPC infrastructures\footnote{In this paper, a federation refers to a set of HPC resources geographically distributed used in collaboration for a workflow execution}.

The paper is organized as follows.  Section~\ref{sec:use_cases} presents use cases where complex workflows integrating different HPC, DA and AI techniques are required to efficiently solve different scientific and industrial problems.
Section~\ref{sec:back_chall} analyses the context of HPC, DA and AI convergence 
as well as the related work, and 
identifies the main challenges toward efficiently supporting these new complex workflows.  Section~\ref{sec:solution} presents a novel approach to address these challenges. 
Section~\ref{sec:discusion} discusses several key decisions of our solution as well as how they compare to alternative approaches.
Section~\ref{sec:conclusion} draws the conclusions from this work and proposes guidelines for future research directions.

\section{Use Cases} 

\label{sec:use_cases}






This section describes three selected use cases from thematic areas, with high industrial and social relevance, that can benefit from innovative and a more holistic workflow approach. These areas target very different users/communities and needs, specifically referring to digital twins in manufacturing (Section \ref{sec:pillarI}),  climate modelling (Section \ref{sec:pillarII}), and urgent computing for natural hazards (Section \ref{sec:pillarIII}).


\subsection{Digital twins in manufacturing}
\label{sec:pillarI}
\begin{figure}[!htb]
    \centering
    \includegraphics[width=1\linewidth]{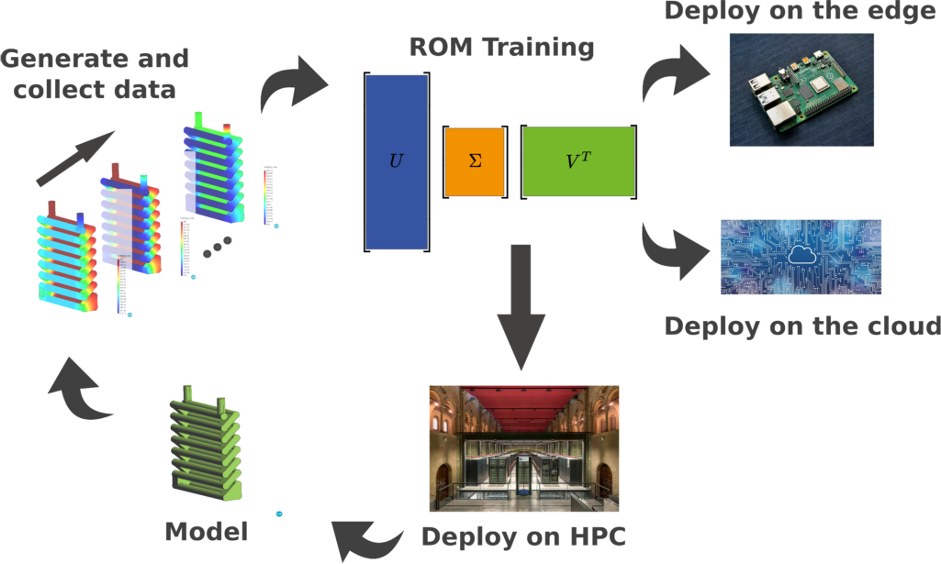}
    \caption{Main phases in the workflow for the construction of ROM models.}
    \label{fig:WP4_general_phases}
\end{figure}

Today, the maturity of numerical methods allows the simulation of realistic problems in manufacturing and the definition of realistic digital counterparts, known as "Digital Twins" of the object or process of interest.
Simulation-based design can nowadays largely substitute experimentation in many fields of application. The predictive value of the numerical models comes however  at the price of a high computational cost. This becomes a blocker in different practical scenarios, and in particular when the objective is deploying the Digital Twin as a companion of the manufactured object for edge computing purposes (for example, on the on-board computer of production machines).
For this application it is necessary that the simulation model provides its results synchronously with the real world, meaning that it has to incorporate input from live sensors and provide immediate results. Currently this requirement can only be fulfilled by extremely simplified models with very limited capabilities. Information on spatial distribution, which would be necessary to identify critical locations such as thermal hot-spots, is out of reach for online applications.  
This limitation can be solved via Model Order Reduction approaches that allow the definition of ``surrogate models'', known as "Reduced Order Models" (ROM), which present a similar predictive value but a much reduced computational cost. 
The essential idea at the basis of such approaches is to perform first a campaign of high fidelity numerical experiments (known as Full Order Models or FOM) in order to collect training data. Such data is then analyzed in search of the most relevant patterns, typically using large-scale Singular Value Decomposition (SVD) techniques.  Finally, the identified patterns are fed back to the original simulation model which exploits them to construct the target ROM model. The corresponding workflow is shown in Figure~\ref{fig:WP4_general_phases}. 

The overall outcome is that the ROM model provides a tunable approximation (that is, an approximation with a controllable level of accuracy with respect to the original FOM model) albeit at a fraction of the computational and memory costs required by the FOM.
These projection-based ROM approaches can be viewed as a special class of machine learning (ML) techniques, characterized by an overall workflow that adheres to the classical training and inference model. From a computational point of view, the training part is particularly challenging since it requires dealing both with the generation of ``experimental'' data and with their analysis via large-scale SVD. 
The generation of training data implies a classical finite element solution, implemented through the software Kratos \cite{Dadvand2010}\cite{Dadvand2013}, with a computational cost governed by the cost of solving linear systems of equations. This step is typically addressed within the project by the use of an algebraic multigrid solver \cite{Demidov2019},\cite{Demidov2020}. 
The computation of the SVD is identified as a computationally critical kernel in the reduction workflow as it forms the basis for various of the reduction algorithms as well as for the hyperreduction steps. The computation of the large scale SVD faces challenges related to both its computational cost and to the memory requirements. For example a workflow including 10 million degrees of freedom (dof) and 5000 time steps would require 400~GB of memory simply to store the unprocessed data, with the memory requirements becoming sensibly higher for any manipulation of such data. While careful out-of-core approaches (disk swap) allow overcoming this memory limitations, they provide a practical bottleneck to any dynamic workflow due to the inherently large increase in the computational time (hours instead of minutes). 
To overcome this challenge, the implementation of a \emph{distributed} randomized truncated SVD able to deal with a large amount of distributed memory while at the same time reducing the computational time is required. 


A complete workflow may also require an iterative refinement of the training campaign to deal with gaps in the training space. The effective use of supercomputers requires integrating both the training and the inference steps within a single complex workflow that is adapted to the specific needs of the problem to be addressed. 
This in practice implies that, while a ``classical'' workflow would require manual iterations between training and evaluation, integration within a workflow management system allows automation of the process with obvious advantages in real applications.

A practical challenge is the need to deploy  the different software stacks on multiple hardware configurations. 
This represents a nontrivial challenge given the strong dependency on system libraries, such as the message passing interface (MPI).

Furthermore, the described workflow can also be integrated with other AI frameworks with the aim of eventually employing the ROMs as building blocks in the construction of system-level models. This integration also poses relevant
challenges, particularly regarding the interoperability between the different modules to be integrated within the same workflow.

\subsection{Climate modelling}\label{sec:pillarII}

The study of climate change and related climate phenomena is extremely challenging and requires access to very high-resolution data. In this respect, the climate community has been continuously pushing the boundaries to deploy and run model simulations at the highest resolution possible, exploiting cutting-edge supercomputing infrastructures \cite{schulthess19}. The resulting output consists of large, complex and heterogeneous datasets that require proper solutions for management and knowledge extraction \cite{elia21}, and which can take advantage of data-oriented approaches from DA and AI fields. 

Typical end-to-end Earth System Modelling (ESM) workflows consist of different steps, including input data pre-proce\-ssing, numerical simulation runs, output data post-processing, as well as DA and visualization. Even though they represent different parts of the same scientific discovery process, their \textit{seamless}, \textit{intelligent} and \textit{efficient} integration into HPC environments still needs to be addressed at a variety of levels to become a reality.

The methodologies currently available for developing scientific workflows in the climate field are unable to 
integrate  the whole set of components transparently into a single workflow
The current approach usually relies on non-standard, home-made scripts to address the operational needs and integrate different components into the ESM workflow. Most often, these efforts focus on a specific HPC machine, which makes it extremely hard to port the same solution to other HPC facilities. To this end, ESM workflows can benefit greatly from enhanced solutions that hide the underlying technical details of HPC machines and provide standardized ways to develop and integrate the distinct ESM workflow components, including DA and AI components. This could lead to improved execution efficiency, along with an optimized usage of HPC resources and increased research productivity.





In this respect, the improvements
provided by dynamic access to the model simulation results at runtime, together with AI techniques, can be exploited as part of the ESM workflow management. They can bring forward advanced possibilities for smart execution of the workflow, enabling more efficient resource usage as well as a shorter time-to-solution. One of the typical tasks in climate modelling is to run ensemble simulations that consist of multiple members and can take a significant amount of time. Ensembles are used to assess uncertainty in model results, for model tuning, or for exploring different scenarios of particular events. 
In the current workflows, the number of members to run is usually fixed at the start of the workflows and is constrained by the available computational resources.
Ultimately, not all the members may be needed. In this sense, dynamic workflows with in-memory access to model results, able to adapt simulations at runtime by performing a smart (possibly AI-driven) pruning of ensemble members, could reduce resource usage and improve energy efficiency. One of our major objectives is to determine which metrics can be used to prune members without impacting the quality of the simulation.  This requires novel systems able to adapt dynamically the workflow execution, based on these runtime-computed metrics. In a more general sense, dynamic access to model results allows the implementation of model diagnostics, especially those that require high temporal frequency data without changing the model code and frequent data serialization. This is especially important for very high resolution climate models, which face I/O and storage limitations.

Data-driven approaches can also play a significant role in enhancing knowledge extraction from large climate simulation data, leading to a better understanding of the climate system. In this respect, Tropical Cyclone (TC) detection and tracking represents an important case study since it requires multiple two-dimensional fields, such as \textit{pressure, temperature, wind velocity, vorticity}, at different time steps (with a frequency of at least 6 hours) and from very high-resolution General Circulation Model (GCM) data \cite{scoccimarro11}, for example coming from the Coupled Model Intercomparison Project - phase 6 (CMIP6) \cite{eyring2016} or very high-resolution models (e.g. the CMCC-CM3 model). 

TC analysis can be very challenging due to the large amount of data involved, its heterogeneity, and processing complexity. This is even more critical if data from multiple models are considered in the process. Different detection and tracking methods, mainly based on statistical methods, are available in the literature \cite{horn2014}, and new emerging approaches are investigating the use of ML/deep learning (DL) techniques to assess the possibility of speeding up the process and improving energy efficiency. Currently, these types of analyses  are generally executed offline, on the data produced by the ESM models, using specialized tools and scripts and could benefit greatly from  novel solutions able to include DA and ML/DL technologies in the HPC workflow. The adoption of strongly integrated and data-driven approaches will enable scientists to tackle considerably larger and more complex problems than are possible today in climate science. In-situ mechanisms will represent another step forward in this direction by integrating data-driven approaches directly within the model simulation, delivering an even more efficient solution. 



\begin{figure}[htb!]
    \centering
    \includegraphics[width=1\linewidth]{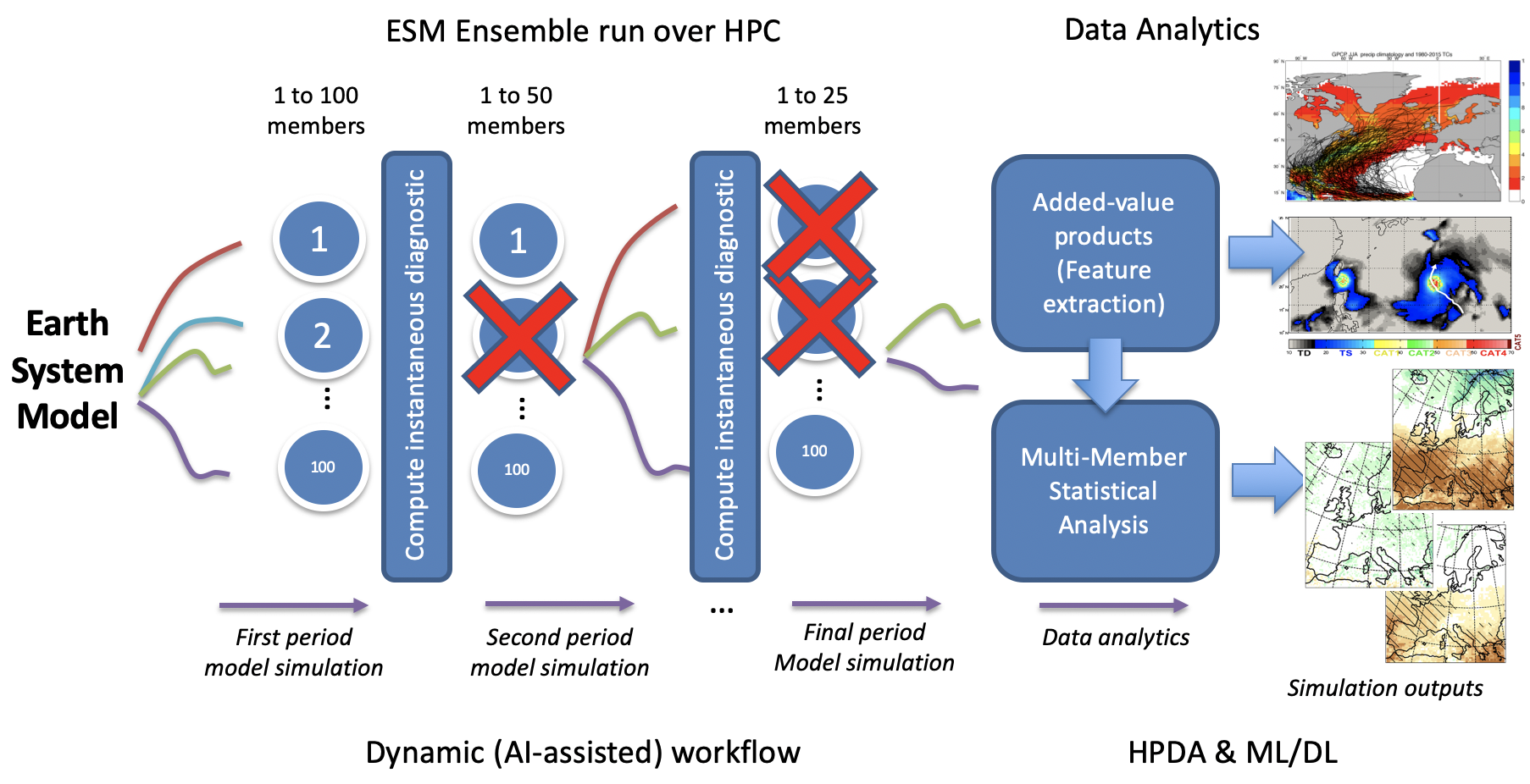}
    \caption{ Main phases envisioned in the enhanced ESM workflow. The left side shows the ensemble members pruning process, while the right side depicts the feature extraction stages based on HPC, DA \& ML/DL techniques.}
    \label{fig:pillarii}
\end{figure}

Figure \ref{fig:pillarii} shows the resulting end-to-end ESM workflow integrating the aforementioned aspects. This can only be possible through new workflow platforms able to integrate HPC, DA and AI components.

The availability of enhanced workflow capabilities will ultimately (i) support transparent integration of simulation-centred and data-driven components, (ii) allow scientists to further increase knowledge of the climate system by delivering better data to end-users for societal challenges, and (iii) democratize access to these complex end-to-end ESM workflows.

\subsection{Urgent computing for natural hazards}
\label{sec:pillarIII}

Urgent Computing (UC) applies HPC/DA during or immediately after emergency situations, combining complex edge-to-end workflows with capacity computing. Early decisions in earthquake/tsunami response are typically based on interpretations of the best available, yet often limited, data immediately following the event to estimate its impact. 
Synthetic maps of ground shaking and/or tsunami inundation can help to assess losses (e.g., for the insurance sector) or to direct immediate relief (e.g., for civil protection and first and second responders). Ensemble realizations are typically required to account for input and model uncertainties, and have strict time constraints (e.g., two hours in ARISTOTLE-eENHSP, enhanced European Natural Hazard Scientific Partnership~\cite{web:aristotle}). Given their inherent computational cost and input data sensitivity, significant HPC resources are needed to enable high-fidelity large-scale ensemble simulations within the required time constraints.

\begin{figure}[htb!]
    \centering
    \includegraphics[width=1.0\linewidth]{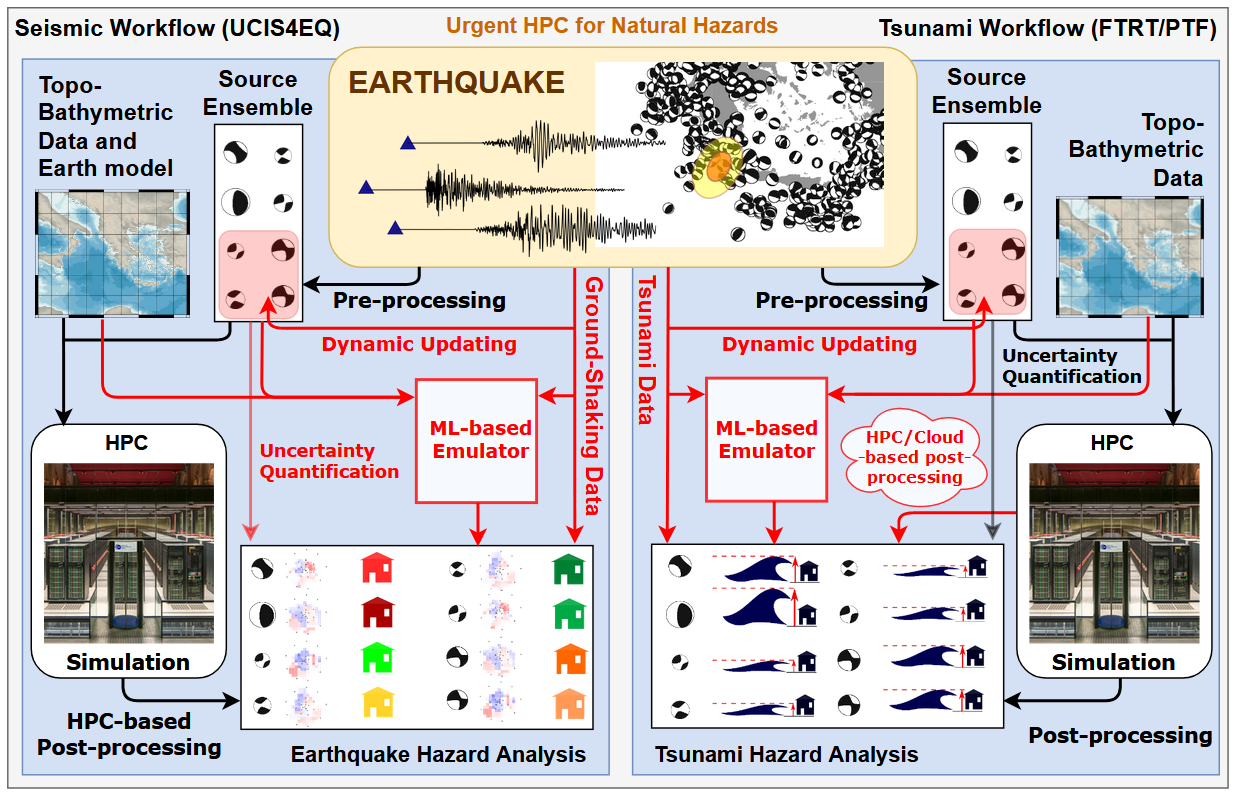}
    \caption{Main phases in the UC workflows for earthquakes and tsunamis. Black elements describe the existing workflows and red elements the desired advances.}
    \label{fig:WP6_general_phases}
\end{figure}

Seismic and tsunami UC workflows consist of three main phases (see Figure \ref{fig:WP6_general_phases}):
\begin{enumerate}
\item Pre-processing, where an ensemble of possible earthquake sources (with uncertainty) is based on seismic data.
\item Simulation, where ground shaking/tsunami impact is quantified numerically for each individual scenario.
\item Post-processing, in which simulation results are aggregated to produce probabilistic estimates including both source and modelling uncertainty, potentially updated with incoming observations from monitoring networks.
\end{enumerate}
Figure \ref{fig:WP6_general_phases} displays the existing workflows and
planned development extensions. In particular, future UC workflows aim at being responsive to live data streams with dynamically updated scenario ensembles and hazard analyses. ML-based emulators may be able to predict outcomes of the HPC-based simulations given sufficient training sets. The tsunami post-processing currently performed on local hardware may be improved by being performed on HPC resources, and a prototype of a database of Earth models will be developed for the seismic workflow.

\subsubsection{Probabilistic tsunami forecasting and faster than real time tsunami simulations}
Probabilistic Tsunami Forecasting (PTF) for tsunami early-warning requires that a large ensemble of tsunami simulations are calculated Faster Than Real Time (FTRT), based on source estimates immediately following an earthquake \cite{Selvaetal2019,Lovholt2019,goubier2020,Giles2021,Selvaetal2021}.  Uncertainties arise both from model limitations and scarce knowledge of fault geometry and mechanism. Both are managed in PTF \cite{Selvaetal2019}. Ensemble initialization is based on real-time seismic monitoring tools and simulations are performed with the Tsunami-HySEA code running on Graphics Processing Units (GPUs). Post-processing aggregates the individual simulations, managing inherent uncertainty. 
The following specific steps may improve the operational level of tsunami forecasting:

\begin{enumerate}
\item  Revision of the PTF workflow to reach time-to-solution performance targets relevant for UC, with time-management and failure safeguards.
\item Optimization of ensemble initialization and updating with dynamically evolving uncertainty quantification, based on real-time seismic data, tsunami records, and DA. 
\item Use of AI for rapid impact estimation (e.g. \cite{salmanidou2017,mulia2020,makinoshima2021}) to accelerate the workflow and potentially integrate ensembles in real-time. 
\item Use of DA and AI tools to enhance event diagnostics and post-processing analyses including cloud storage of detailed results for subsequent DA processes.

\end{enumerate}

To meet time constraints, the workflow will benefit from a solution that overlaps multiple phases, avoids global synchronizations, and exploits workflow environments able to perform dynamic elastic resource management. Fault tolerance is an additional requirement, as is an environment able to integrate HPC phases with DA and AI. 

\subsubsection{UCIS4EQ}

The Urgent Computing Integrated Services for Earthquakes (UCIS4EQ) workflow has been developed as a Pilot Demonstrator under the ChEESE Center of Excellence\footnote{https://cheese-coe.eu}. UCIS4EQ coordinates the execution of large 3D full waveform simulations to obtain rapid and realistic synthetic shaking estimates following an earthquake \cite{DelaPuente2020}. UCIS4EQ is coupled to state-of-the-art massively parallel simulation solvers so that, given sufficient HPC resources, simulations can be completed within minutes to hours. Typical uncertainties come both from source characteristics that cannot be constrained uniquely (given sparse initial data) and from unresolved soil effects which may amplify or reduce seismic waves. 

UCIS4EQ is developed considering not only the functional requirements, but also to ensure the quality of non-functional requirements such as robustness, interoperability, availability, portability, and maintainability. Each process is encapsulated to work as a specialized micro-service, with all components containerized and ready to deploy as a cloud service. The following specific aspects should be upgraded in the UCIS4EQ workflow (under development or not implemented) to bring it closer to an operational level:

\begin{enumerate}
\item An integrated workflow management system, including workflow monitoring and steering and dynamic resource management. Currently the components are activated sequentially, with HPC jobs submitted simultaneously.
\item Ensemble simulations for uncertainty quantification. A current prototype exists for suites of runs without assigned probabilities.
\item Real-time data assimilation to assess and/or adjust the ongoing simulations. 
\item A prototype database of 3D velocity models. 
\item Regression and/or DL models to estimate shaking intensity maps at time scales of seconds to minutes. Such ML ``twins'' will enhance uncertainty quantification. 
\end{enumerate}

To achieve the use case objectives, the workflow will benefit from a solution that enables autonomous integrated management of the multiple HPC and  data assimilation phases, with support for a real-time assessment of the  situation to enable a dynamic adaptation of the simulations.

    \section{Background and challenges}
\label{sec:back_chall}

This section analyses the context of HPC/DA/AI convergence for supporting the new complex workflows presented in the previous section. First, we present background and previous work from several points of view (development, deployment, data management and computer architecture). Afterwards, we present the main challenges to efficiently supporting these complex workflows.

\subsection{Background and related work}
\label{sec:background}





\subsubsection{Workflows development and HPC}


An important aspect in scientific workflows concerns the programming structures they provide for the development of scientific applications~\cite{talia2013sworkflow}.  
Existing approaches in this area can be broadly categorized based on their level of abstraction model (high-level versus low-level) and on the type of programming formalism they support; some of them are based on graphical interfaces, such as Kepler~\cite{altintas2004kepler}, Taverna~\cite{wolstencroft2013taverna} or Galaxy~\cite{galaxy}; some on textual interfaces, such as Pegasus~\cite{deelman2005pegasus}, Askalon~\cite{fahringer2005askalon} or Autosubmit~\cite{autosubmit}; and several on programmatic interfaces, such as COMPSs~\cite{compss_servicess} or Swift~\cite{wilde2011swift}.

A relevant observation is that  scientific communities seem to stick to one or another solution. For example, Galaxy~\cite{goecks2010galaxy} has been adopted by the ELIXIR  life science research infrastructure as its main workflow environment, while Cylc~\cite{oliver2018cylc} was selected, among others, by the Earth Science community.

Typically, HPC applications are developed using the MPI programming model~\cite{gropp1999using}, which is the de-facto standard for this type of applications. It is based on the idea of having a large number of concurrent processes exchanging messages to solve a large problem cooperatively. MPI is combined with other approaches to exploit concurrency inside the potent HPC nodes. The most popular approach for this is OpenMP~\cite{dagum1998openmp}. Additional complexity for the application developers is the appearance of accelerators, such as GPUs, which require specific programming environments such as CUDA~\cite{nvidia2007compute} or OpenCL, among others. HPC programming models tend to be quite low level and require considerable effort from the application developer. 


\subsubsection{Data analysis workflows}
DA applications can be conveniently modelled as workflows combining distributed datasets, pre-processing tools, data mining and ML algorithms, and knowledge models. 
Compute and storage facilities of large-scale HPC systems can be effectively exploited to parallelize the execution of workflows composed of dozens to thousands of tasks, to achieve higher throughput and to reduce turnaround times. This is particularly true in the context of DA workflows, in which the data volumes to be analyzed are huge, and tasks take a long time to complete their execution on conventional machines \cite{Marozzo2018}. Implementing efficient DA workflows from scratch on HPC systems is not trivial and requires expertise in parallel and distributed programming.
To cope with design and programming issues, high-performance programming models for data analysis workflows have been recently proposed \cite{DaCostaSFI2015}.

DA workflows allow programmers to express parallelism at several levels (i.e., data, task, pipeline parallelism), which can be exploited at runtime by HPC platforms comprising a large number of processing and storage elements. In addition,
the ability to reuse workflows by modifying the input data or the algorithms and tools, combined with the ability to create hierarchical workflows where individual nodes can, in turn, be workflows, allow users to define and execute a variety of data analysis applications that go well beyond the classical scientific applications executed on HPC platforms.

\subsubsection{AI workflows}
The convergence of AI environments --and more specifically ML libraries-- with HPC platforms provide the opportunity for major performance improvements for the effectiveness, reusability and reproducibility of the simulation \cite{jha2019MLHPC}. Usually, models are generated by running effective ML algorithms over large data sets that are produced from various sources. The generated model can comprise vectors of coefficients as well as different tree or graph structures with specific values. These derived models can accelerate the development of high-performance DL inference applications. Furthermore, pre-trained models also speed up the production deployment process. 

Having a model repository enables the tracking of parameters and results of trained models to further package with ML libraries and codes in a reproducible and reusable manner in a targeted environment. For HPC and DA convergence, storing, managing, and sharing capabilities of models are key requirements for building workflows that make use of ML/DL models. 
Some efforts towards automatic management of ML in HPC systems are Dkube\footnote{URL{https://www.dkube.io/products/datascience/hpc-slurm.php) }} or the CODEX AI suite\footnote{URL{https://atos.net/en/solutions/codex-ai-suite-fast-track-artificial-intelligence}}



\subsubsection{Deployment in large infrastructure systems}
\label{sec:context_usage}

Due to the widespread number of compilers, library versions, and their incompatibilities, every time users want to deploy a new workflow in a supercomputer, they have to check the dependencies, and install the missing ones taking into account the libraries and compiler versions to detect possible incompatibilities. 
To mitigate these issues, there exist tools such as Spack~\cite{spack} or Easybuild~\cite{easybuild} that provide mechanisms to deal with these issues and automate the installation process of new software in HPC environments. However, they still require an expert HPC developer to create the packages or recipes for these tools and verify that they work for each supercomputer.


In Cloud environments, virtualization and container technologies have simplified the portability of complex applications. Hypervisors such as KVM~\cite{kvm} or container engines such as Docker~\cite{docker} allow running processes in customized environments on top of computing nodes. These environments can be customized as normal computers and saved as images, which can easily be copied to other nodes where the same process can be executed with the same environment. 
The main barriers to deploying these technologies in HPC are rooted in the requirement of running hypervisors and engines in privilege mode (root access) with the security consequences that this implies and the integration of images with specific HPC hardware such as fast interconnects drivers. 
Singularity~\cite{singularity} is a container engine that tries to overcome these issues by not requiring privileged user mode to run the container, allowing direct access to the host drivers to benefit from the special HPC hardware. Cloud Computing also provides service-oriented abstractions called Everything-as-a-Service, where a set of services is offered depending on the usage model. One of the latest proposed service models is Function-as-a-Service (FaaS). This service enables users to execute functions in the Cloud in a transparent way with a simple REST application programming interface (API) call and without having to deal with the entire deployment, configuration and execution management overhead. The FaaS platforms, such as the commercial AWS Lambda, Google Cloud Functions, or open-source approaches like OpenWhisk~\cite{openwhisk} or OpenFaaS~\cite{openfaas}, are in charge of managing the different function executions, allocating the computing resources when required, deploying the function software, obtaining the input data and storing the output results. 

\subsubsection{Data management/storage aspects}

Persistent storage in HPC has traditionally been dominated by file systems \cite{storage}. Applications consuming file-based data need to open and read the files to load the data into memory, transforming it to the appropriate data structures for efficient manipulation. This process is usually implemented by the programmer as part of the application unless using specific file formats, such as NetCDF \cite{netCDF} or HDF5 \cite{hdf5}, which provide specific libraries to facilitate this task. Additionally, scalability problems in file-based storage systems are well-known \cite{storage}, which led to different storage solutions based on abstractions other than files (e.g., object stores or key-value stores, among others) gaining popularity not only in Cloud but also in HPC environments \cite{daos}. These storage solutions can provide more flexibility in accessing persistent data by enabling data accesses at a finer granularity, as well as providing efficient access to data during the computation, and facilitating the implementation of common application patterns in HPC, DA and AI, such as producer-consumer or in-situ analysis or visualization.

New technologies blurring the line between memory and storage have recently become available. These technologies, called persistent memories or non-volatile memories (NVM), such as Intel Optane DC \cite{optaneDC}, are similar to memory in speed, similar to disk in capacity, and are byte-addressable. These features open the door to computing directly on the stored data without having to bring it to memory, enabling HPC, DA, and AI applications to deal with larger volumes of data (i.e., not fitting in main memory) at high-speed \cite{hotStorage}.


\subsubsection{Computer architecture perspective}
The recent trends in HPC confirm that a hybrid architecture combining CPUs, GPUs, and even customized accelerators has become the preferred node type for a large range of workloads of interest for HPC and data centres, including ML, DA, and scientific simulation. The path to keep increasing performance while maintaining energy efficiency lies in the use of Domain-Specific Architectures (DSAs). Indeed, one of the most prominent and appealing domains for specialization and adaptation of the system is DL. New accelerator units such as Google's TPU \cite{web:tpu} offer considerable higher energy efficiency when compared with traditional architectures (CPUs or even GPUs). Adoption of such DSAs to workflows becomes significant. In addition, over the last years, reconfigurable devices, such as FPGAs (field-programmable gate arrays), are gaining popularity as co-processing devices in HPC and Data centre environments. There are clear past successful examples such as the Catapult project \cite{putnam2014catapult}. Moreover, new products based on FPGAs, such as Alveo and Versal boards, target AI applications and the HPC domain.


\subsubsection{Other related projects}
\label{sec:related_work}



We can find several EU funded projects targeting the convergence between HPC and DA. 
The LEXIS project \cite{web:lexis,lexis2020,lexis2021} tackles the description and automation of computational workflows (e.g., simulations with subsequent data analysis steps) providing easy access to federated computing and data systems. 
It is building an advanced engineering platform and portal leveraging large-scale geographically distributed HPC and Cloud computing resources. 



Similarly, the EVOLVE project~\cite{web:evolve} also proposes a software stack to integrate HPC and DA environments. It consists of a Python module for Apache Zeppellin notebook, which is used to create workflows based on container images deployed and orchestrated on top of a Kubernetes architecture. It allows to transparently deploy applications both in the Cloud and in clusters. 


The ACROSS project~\cite{web:accross} is a EU funded project that aims at combining traditional large scale HPC simulations, high-per-formance data analytics (HPDA) and ML/DL techniques to boost the performance of the simulation frameworks and/or improve the quality of the simulation results without increasing computing resources consumption.

The ADMIRE project 
\cite{web:admire} focuses on satisfying the performance requirements of today's data processing applications by addressing the storage bottlenecks in HPC architectures. To this aim, the project proposes to create an active I/O stack that dynamically adjusts storage resources to the computational resources of jobs, taking advantage of emerging multi-tier storage hierarchies including persistent memory devices. 




\subsection{Challenges} 
\label{sec:challenges}
The context and requirements of the described use cases have raised a set of challenges in workflow design and management that are summarised below.

\subsubsection{Challenge 1: Enable the openness, reusability, reproducibility and accessibility of the workflows and their results}
With the introduction of virtualization and containers, porting applications into Cloud environments has improved considerably, easing the porting of applications implemented in well-known software stacks (e.g., LAMP, MEAN, Hadoop). 
However, the complexity of workflows is growing fast, and they are required to combine multiple HPC, DA, and AI frameworks. Enabling the portability and accessibility of these workflows to the wide variety of HPC systems is still an open challenge. 
First, current workflows require deploying and orchestrating the multiple frameworks, which must be coupled tightly with the computing infrastructure. Moreover, reusing the same tools used in the Cloud environments to deploy applications in HPC environments is not possible due to the security and accessibility requirements in supercomputers. Therefore, installations and deployments are usually managed by system administrators to ensure they are adapted to the supercomputer capabilities.  


A similar challenge appears with the workflows results. Enabling their reusability and reproducibility requires designing and implementing a tailored mechanism to make those results available to the users considering the access restrictions of the HPC systems.
To overcome this situation, new tools or current Cloud deployment and data-sharing tools have to be redesigned, extended, and adapted to accommodate the requirements of the new complex workflows and to fulfil the HPC access constraints. 
This includes related to user management, security and version control of the workflow and their components (data, model, code components) in federated infrastructures.



\subsubsection{Challenge 2: Simplify the development of complex workflows while keeping their capabilities and performance}


Traditionally, the HPC software stack has focused on providing libraries to optimally exploit the target infrastructure. As mentioned earlier, existing HPC programming models, such as MPI or OpenMP, enable the development of parallel applications but they are still too complex for general scientists, especially if they need to develop a higher abstraction, complex, workflow.   
Even more, existing workflow systems usually do not entail the possibility of including parallel or HPC tasks (i.e., tasks implemented with MPI and/or OpenMP).


Also, as mentioned in section~\ref{sec:introduction}, different methodologies have been proposed for the development of HPC, DA, and AI codes. 
In general, current available methodologies for developing workflows do not fulfill the requirements of increasingly complex applications,  requiring novel procedures supporting a holistic workflow composed of HPC simulations or modeling, DA and AI tasks.
To keep performance, the usage of new, powerful, and energy-efficient heterogeneous computing nodes is a must. For example, GPUs are very efficient in the DL training phase, but they require the development of specific code for heterogeneous devices. 




To address these issues, new methodologies that support simpler and intuitive workflow development need to be proposed. Since our focus is to support workflows that integrate components of diverse nature (HPC, AI, and DA), these methodologies should bridge the gap between the application and the heterogeneous infrastructure in order to maintain the expected performance.


\subsubsection{Challenge 3: Support for workflow dynamicity}
An additional challenge introduced by complex workflows is dynamicity. Current workflow managers support static workflows with very limited dynamicity. This approach fits traditional workflows that solve problems with simple pipelines or graphs, repeated loops with different input parameters, or small workflow modifications using conditional control flows.  However, 
the complex workflows considered in this work require support for high degrees of dynamicity and flexibility in their development and execution. 
The workflow programming models and engines should support applications with dynamic data sources, with variable input data, producing alterations in the computation workflow, invalidating the initiated computations and requesting new computations. In some cases, the workflow requires a dynamic adaptation applied in real-time to fulfill an urgency constraint of the computation. 
To support this dynamicity, the workflow manager should be able to react to changes in  the input data and generate new computations on demand. 

However, the workflow dynamicity  can not only be driven by the input data.  For example, an early analysis of the results can detect parts of the workflow tending to solutions that are insignificant for the final results.  Such computations can be canceled for saving resources or dedicating them to extend the search space or to increase the effectiveness of the solution. In this sense, the workflow manager should be also able to dynamically modify an existing workflow by removing already expected computations, and sometimes adding new ones in reaction to given events. 

Finally, the mentioned dynamic workflow support at the development and execution level has to be tightly combined with elastic resource management in order to adapt the computing capacity with the changing computing demands required by the workflows, which will make more sensible use of the resources and save power.  

\subsubsection{Challenge 4: Enable data management and computation integration}
Current approaches to implement data management usually differ between environments that execute scientific applications and those that implement DA. However, workflows integrating both kinds of computation can improve the productivity of scientists and engineers, as well as the performance of the workflows themselves, which could start the analysis of the data before the data generation ends. To provide an effective environment for this kind of hybrid workflows, it is necessary to provide a unique data management strategy that reduces the data movements between different storage systems, whilst supporting both scientific and DA workloads efficiently. This unique strategy should be able to provide the data generation application with a fast data ingestion mechanism, which should not limit the potential parallelism of the computation (avoiding synchronisation points due to data storing). At the same time, it should be able to provide a simple interface that enables programmers to access intermediate results efficiently.

Datastores for DA usually meet these requirements. However, developers of HPC applications, used to working with files, are reluctant to adopt them for several reasons. First, the efficient utilisation of this type of datastores involves a low-level knowledge of their design, to tune all the available configuration parameters. Second, deciding how to organize the data (i.e., defining the data model) influences the performance of both reading and writing, and obtaining efficient data models also requires a deep knowledge of the execution platform (both hardware and software stack). Third, enhancing data locality is also a goal of this type of data store, but once again it is rarely transparent to programmers. Finally, changing their traditional approach to storing data involves learning new interfaces that usually change from one datastore to another. 

To overcome this reluctance, it is necessary to add a layer to the software stack that relieves programmers of these tasks. This layer should provide automatic and transparent tuning of the data store and data locality enhancement, automatic data modelling, and a simple interface independent of the particular data store in the system and close to the data structures managed by the application. These features would allow the programmer to focus on the problem domain, and at the same time, provide the required performance and parallelism in collaboration with the programming model.

In addition, the popularisation of persistent memory devices offers the possibility of rethinking strategies both on how data is accessed and how data is modelled in datastores. By exploiting the capabilities of these devices in the data management layer, again transparently to the programmer, applications will be able to seamlessly manage larger amounts of data and benefit from a higher performance in data access.

\section{eFlows4HPC solution}\label{sec:solution} 
\begin{figure}[htb]
    \centering
    \includegraphics[width=1\linewidth]{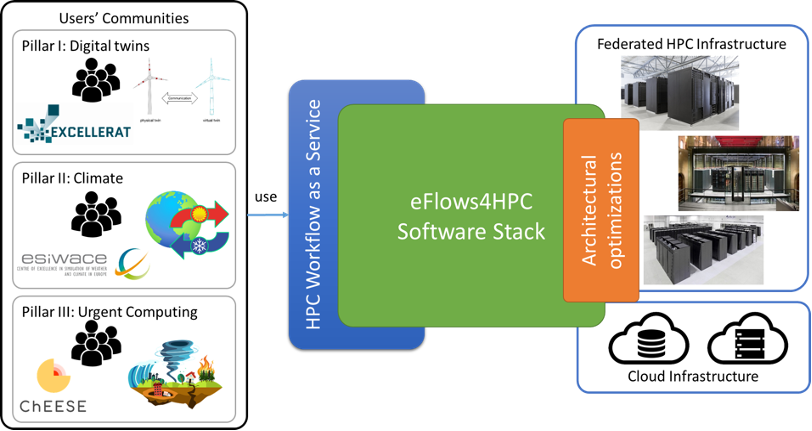}
    \caption{eFlows4HPC project overall approach.}
    \label{fig:approach}
\end{figure}

eFlows4HPC is a EuroHPC funded project which aims at enabling dynamic and intelligent workflows
in the future European HPC ecosystem. The high-level structure of the project is depicted in Figure~\ref{fig:approach}. We propose integrated solutions to cover the challenges presented in Section~\ref{sec:challenges}. First, eFlows4HPC defines a software stack that covers the different functionalities to support the whole lifecycle of the complex workflows introduced in this paper. Second, it proposes the HPC Workflow-as-a-Service (HPCWaaS) methodology to enable reusability of these complex workflows as well as simplifying the accessibility to HPC resources. Finally, the project also works on the workflow kernels for new heterogeneous architectures.

\subsection{eFlows4HPC software stack}

\begin{figure}[htb]
    \centering
    \includegraphics[width=1\linewidth]{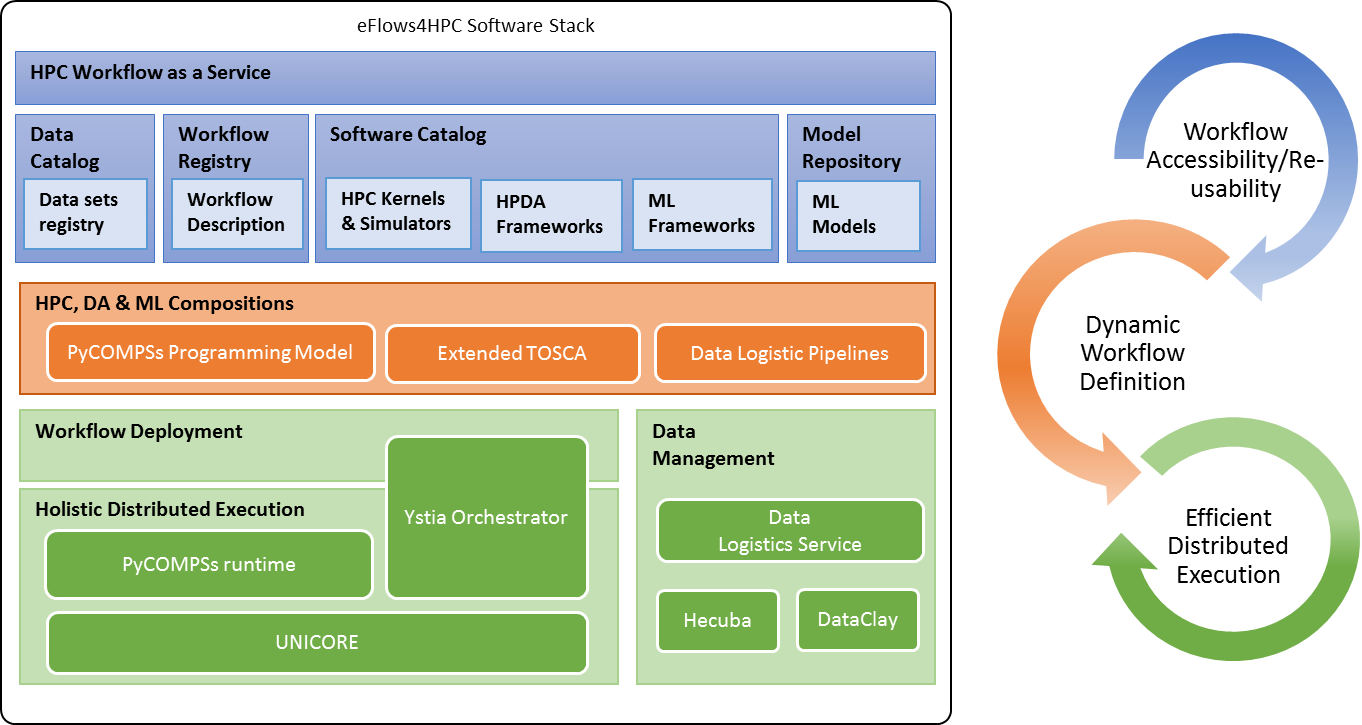}
    \caption{eFlows4HPC Software Stack.}
    \label{fig:stack}
\end{figure}

The eFlows4HPC software stack  comprises existing software components, integrated and organised in different layers (Figure~\ref{fig:stack}). The first layer consists of a set of open repositories, catalogues, and registries to store the information required to facilitate the re-usability of the target workflows (Workflow Registry), their core software components such as HPC libraries and DA/AI frameworks (Software Catalog), and its data sources and results such as ML models (Data Catalog and Model Repository). 
Besides, it also provides the HPC Workflow-as-a-Service interface, which allows developers to  deploy workflows in the HPC infrastructures transparently and makes them easily accessible for the final users. This layer mainly addresses Challenge 1, and the details about how the components are used to address it are described in Section~\ref{sec:HPCWaaS}

The second layer provides the syntax and programming models to implement these complex workflows combining HPC simulations with DA and AI. A workflow implementation consists of three main parts: a description about how the software components are deployed in the infrastructure (provided by an extended TOSCA definition~\cite{tosca}); the functional programming of the parallel workflow (provided by the PyCOMPSs programming model~\cite{pycompss}); and data logistic pipelines to describe data movement to ensure the information is available in the computing infrastructure when required. 
The combination of these three models enables the reproducibility and reusability aspects of Challenge 1 and focuses on the development aspects of Challenges 2 and 3. More details about workflow development are provided in Section~\ref{sec:development}
Finally, the lowest layers provide the functionalities to deploy and execute the workflow based on the provided workflow description. On the one side, this layer provides the components to orchestrate the deployment and coordinated execution of the workflow components in federated computing infrastructures. On the other side, it provides a set of components to manage and simplify the integration of large volumes of data from different sources and locations with the workflow execution. 
This part of the stack addresses the runtime aspects of Challenges 2, 3 and 4. More details on how these components interact at execution time are given in Section~\ref{sec:execution}

\subsection{HPC Workflows-as-a-Service (HPCWaaS)}
\label{sec:HPCWaaS}
Currently, one of the main barriers to the adoption of HPC is the complexity of deploying and executing workflows in federated HPC environments. Usually, users are required to perform software installations in complex systems which are well beyond their technical skills. Therefore, preparing the workflows for execution in a supercomputer typically takes a large amount of time and human resources. If it needs to be replicated on several clusters for reliability requirements or in order to assess the reproducibility of the results, the required time and resources will increase. To widen the access to HPC to newcomers and, in general, to simplify the deployment and execution of complex workflows in HPC systems, eFlows4HPC proposes a mechanism to offer HPC Workflows-as-a-Service (HPCWaaS) following a similar concept as the Function-as-a-Service (FaaS) in the Cloud, but customizing it to workflows in federated HPC environments. The goal is to hide all the HPC deployment and execution complexity to end-users in such a way that executing a workflow only requires a simple REST web-service~\cite{rest} call. 

\begin{figure}[htb]
    \centering
    \includegraphics[width=0.9\linewidth]{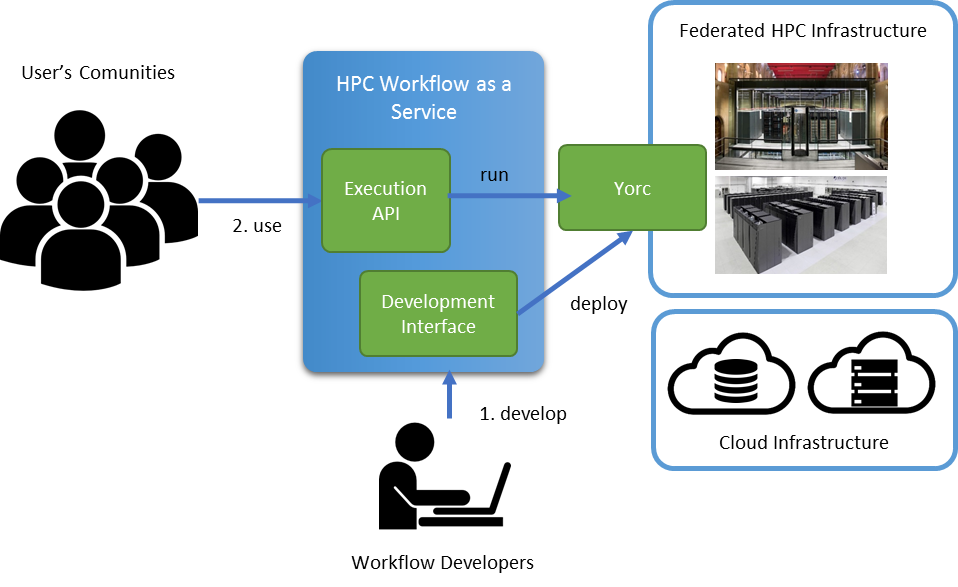}
    \caption{Overview of the proposed HPC Workflow-as-a-Service model.}
    \label{fig:figureHPCWaaSarchitecture}
\end{figure}

Figure \ref{fig:figureHPCWaaSarchitecture} shows an overview of how the proposed model works. The HPCWaaS is built on top of the eFlows4HPC software stack in order to provide the required functionality to develop, deploy, and execute complex workflows. 
Similar to the FaaS model, two user roles are defined in the HPCWaaS: one for workflow developers that are in charge of implementing the workflows, and another for workflow user communities, which are solely interested in executing the workflow and obtaining the results. To support these two roles, the HPCWaaS platform provides two interfaces: the Developer Interface, to build and register the workflows in the system, and the Execution API which provides a REST interface to manage the execution of the registered workflows. 

At development time, workflow developers are in charge of building the workflow using the information stored in the registries and the programming models provided by the eFlows4HPC stack (to be described in Section~\ref{sec:development}). Once the workflow implementation is completed, the workflow is registered in the HPCWaaS platform to make it available to the community of workflow users. On one side, the workflow description will be stored in the Workflow Registry, the description of software components will be stored in the Software Catalog, and the data sources and outcomes will be registered in the Data Catalog and Model Repository. Upon a successful registration, the workflow developer receives a service endpoint from the Execution API, which can be shared with the workflow users to invoke the developed workflow. At invocation, the workflow will be automatically deployed and executed in the computing infrastructure using the rest of eFlows4HPC stack functionalities.

The proposed HPCWaaS model addresses the reusability, reproducibility and accessibility of Challenge 1. Once a workflow is registered to the HPCWaaS, different  workflow users can easily execute the workflow by invoking the workflow end-point that facilitates the accessibility. Several users can run the workflow with the same input parameters that will perform the same computation providing the reproducibility capability in the developed workflows, but they can also invoke the workflow with different input data to perform the same computation on different datasets that also simplifies  reusability. Since the workflow and the descriptions of their software components are stored in public repositories, they can also be re-used to compose other complex workflows.

The following sections provides additional details on how the eFlows4HPC components interact to provide the required functionality in the phases of the workflow lifecycle.

\subsection{Workflow development phase}
\label{sec:development}

\begin{figure}[htb]
    \centering
    \includegraphics[width=0.9\linewidth]{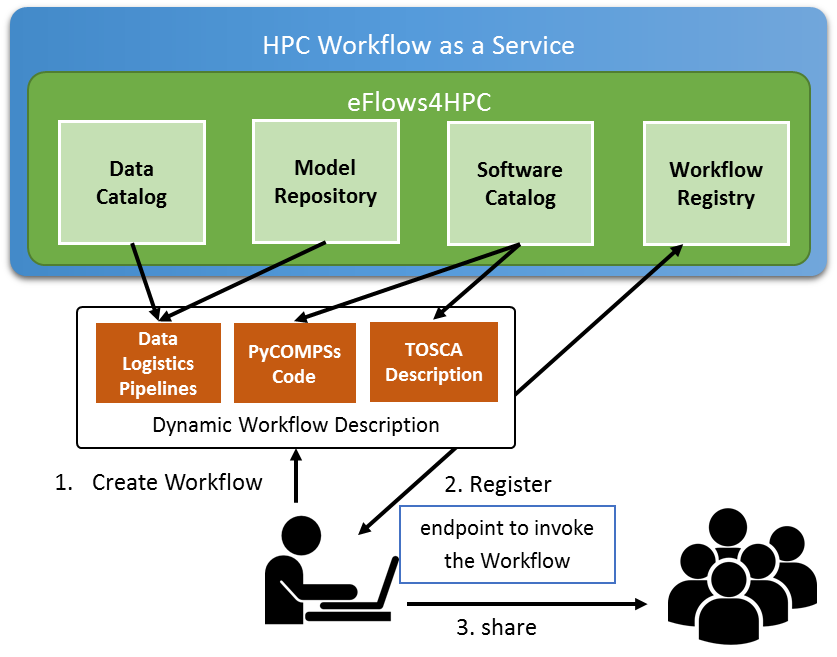}
    \caption{Workflow development phase.}
    \label{fig:workflow_development}
\end{figure}

One key part of the mentioned challenges is the implementation of complex workflows that combine HPC, DA, and AI frameworks in a dynamic and reusable way. eFlows4HPC proposes two mechanisms in order to achieve this challenge, as depicted in Figure~\ref{fig:workflow_development}. 
On the one hand, the software stack provides a set of open catalogs and repositories, providing workflow developers with a means to store and share information about the software components, data and models used by the workflows.  These component are built on top of version control repositories, such as Git, providing the capabilities to manage different versions of the stored descriptions and codes:

\begin{enumerate}
\item The Data Catalog and the Model Repository store the description of those datasets and ML models that are available as input for the workflows or those that are generated by the workflows. They include information on how to access or store them, such as the format, location, and protocol (FTP, WebDAV, etc.). This description is used by the Data Logistic Pipelines to identify the input and output data and how to process them.
\item Similarly, the Software Catalog stores the description of the software components used by the workflows. It stores the information about how the software is deployed and how it is invoked. The first part is included in the workflow TOSCA description and the invocation description is used by the PyCOMPSs workflow to simplify the integration of software executions inside a workflow.
\item Finally, the workflow registry stores the workflow descriptions, which can be retrieved by other users to reproduce the same workflow in other environments or use them as templates to create new workflows.
\end{enumerate}
In addition, we propose a description that uses this information to create complex workflows that combine different software (HPC, DA, and AI frameworks) which are portable and reproducible. This description is composed of a combination of an Extended TOSCA syntax, the PyCOMPSs programming model, and the data logistic pipelines. 

TOSCA is an OASIS standard to describe the deployment topology and orchestration of cloud applications. This standard allows developers to specify the software components and services required by an application, and the relationships between them. For each component, TOSCA describes how it is deployed, configured, started, stopped and deleted. According to their relationships, TOSCA orchestrators, such as Yorc, generate a set of workflows to orchestrate the whole application lifecycle (deployment, execution, and release). In eFlows4HPC, the TOSCA definition is extended to support the deployment and execution of workflows implemented with PyCOMPSs and data logistic pipelines.

The PyCOMPSs programming model provides the logic of how the different software invocations are performed. PyCOMPSs is a task-based programming model that enables the development of workflows that are executed in parallel on distributed computing platforms. It is based on sequential Python scripts, offering the programmer the illusion of a single shared memory and storage space. While the PyCOMPSs task-orchestration code needs to be written in Python, it supports different types of tasks, such as Python methods, external binaries, possibly multi-threaded (internally parallelised with alternative programming models such as OpenMP or pthreads), or multi-node (MPI applications). Thanks to the use of Python as the programming language, PyCOMPSs supports almost all the dynamicity a programming language offers  to developers (loops, conditionals, exceptions) and naturally integrates well with DA and ML libraries, most of them offering a Python interface. 

Finally, in the last part of the workflow description, the Data Logistic Pipelines allow developers to describe how the workflow data is acquired, moved and stored during the workflow life-cycle in order to ensure the data is available in the computing infrastructure when required. The pipelines are also defined in Python, which reduces the entry barrier for the development.

\begin{figure}[htb]
    \centering
    \includegraphics[width=1\linewidth]{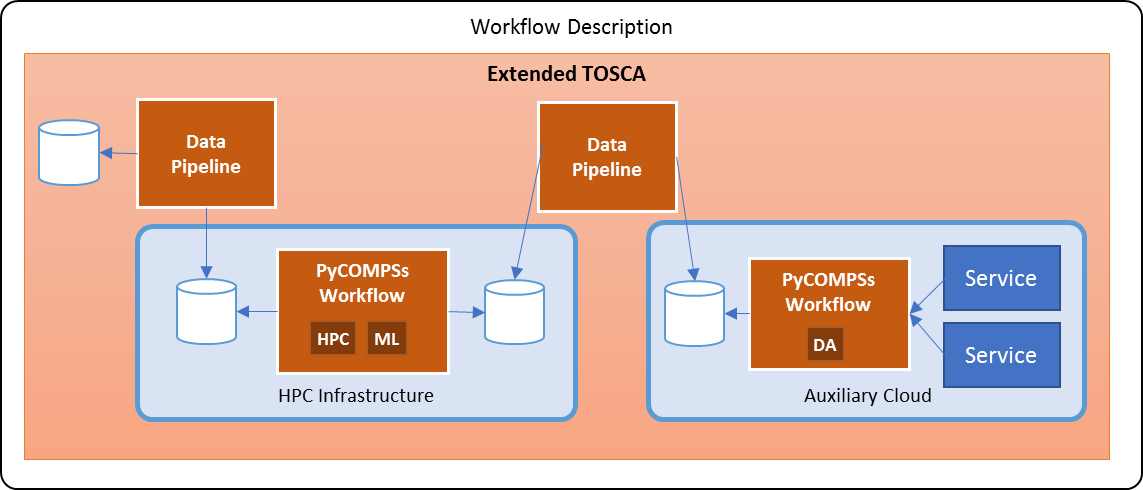}
    \caption{Workflow description example.}
    \label{fig:workflow_example}
\end{figure}

Figure~\ref{fig:workflow_example} illustrates an example of a workflow description. As mentioned earlier, it is mainly an extended TOSCA topology defining the components of the workflow and their relationships. For instance, this example is composed of two PyCOMPSs workflows, two data pipelines, and several services. One of the PyCOMPSs workflows must be deployed in an HPC infrastructure and requires an HPC simulator as well as an ML framework; the other runs in the cloud and requires a DA framework. Regarding the data logistic pipelines, one is defined to retrieve data from an external repository and  move it to a shared storage of an HPC cluster; the other performs movements between the HPC and Cloud environments. The TOSCA description of the workflow components provides a link to specify where the code for the components is stored, and describes how it can be deployed and executed. 

As mentioned earlier, the workflow description is registered and stored in a workflow registry by means of the Development Interface. The result of this registration produces a new service endpoint in the Execution API that can be later used to invoke the execution of the workflow.

\subsection{Workflow invocation phase}
\label{sec:execution}
\begin{figure}[htb!]
   
   \centering
     \begin{subfigure}[b]{1\linewidth}
         \centering
         \includegraphics[width=1\linewidth]{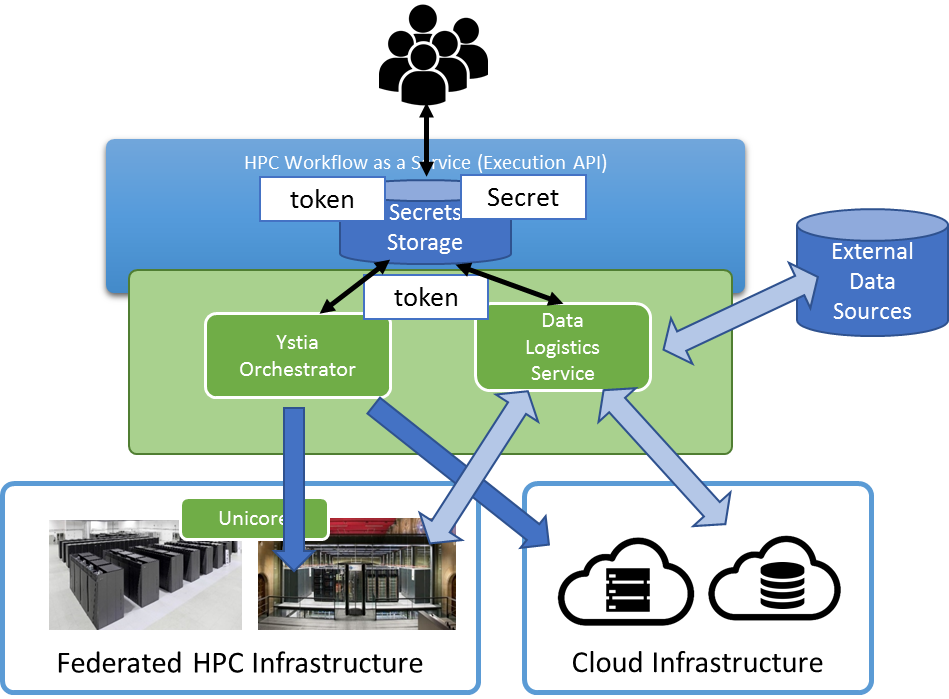}
         \caption{Credentials Management.}
         \label{fig:credentials}
     \end{subfigure}
     \begin{subfigure}[b]{1\linewidth}
         \centering
         \includegraphics[width=1\linewidth]{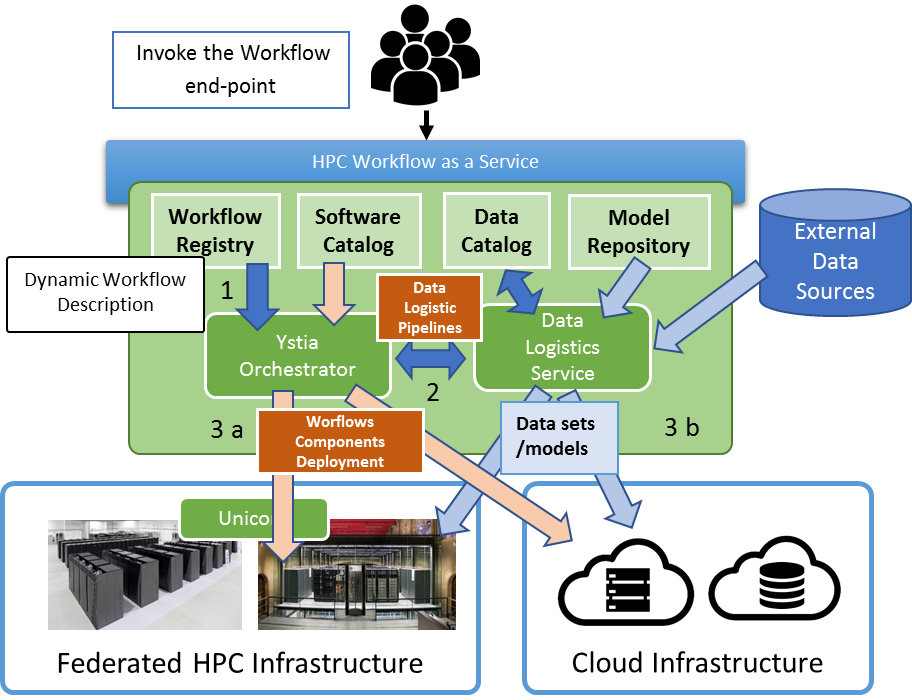}
         \caption{Workflow deployment.}
         \label{fig:workflow_deployment}
     \end{subfigure}
     \begin{subfigure}[b]{1\linewidth}
         \centering
         \includegraphics[width=1\linewidth]{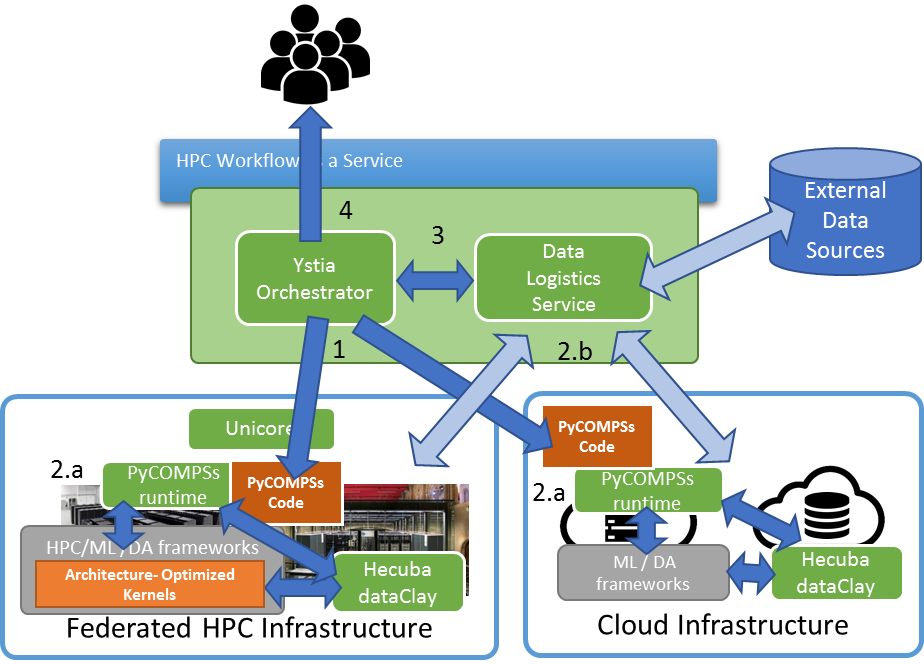}
         \caption{Workflow execution.}
         \label{fig:workflow_execution}
     \end{subfigure}
   
\caption{Workflows invocation phase. Dark blue arrows represent eFlows4HPC component interactions; light blue ones represent data flows; and orange arrows represent component deployments.}
\label{fig:workflow_invocation}
\end{figure}

Prior to executing the registered workflows, the users have to configure the infrastructure access credentials. These consist in the usernames and secrets such as public-key certificates, passwords, etc. The users' certificates are managed by an Execution API, as shown in Figure~\ref{fig:credentials}. This provides a few methods to register and access credentials or generate a new secret, such as a key-pair that the user has to authorize by adding them in the authorized keys of the HPC cluster. The access credentials are stored in a secrets' storage such as Vault~\cite{vault}. These credentials will be identified by a token attached to the user's workflow invocation. This will allow the components involved in the execution to use these secrets to access the infrastructure on behalf of the user, deploy the required components and data, and spawn the workflow computations. 

Once the credentials are registered, the user only need to invoke the endpoint provided at the end of the workflow development phase. As a result of this invocation, the deployment and execution in federated computing HPC infrastructures is triggered. This functionality is provided, as depicted in Figure~\ref{fig:workflow_invocation}, by the cooperation of several components at different levels. At the highest level, the Ystia Orchestrator (Yorc) is in charge of managing the overall workflow deployment and execution. First, it retrieves the workflow description (Step 1 in Figure~\ref{fig:workflow_deployment}) and passes the data logistic pipelines to the Data Logistic Service (Step 2 in Figure~\ref{fig:workflow_deployment}) to set up the required data movements such as the data stage-in and stage-out, or periodical transfers to synchronize data produced outside the HPC systems (Step 3b in Figure~\ref{fig:workflow_deployment} and Step 2b in Figure~\ref{fig:workflow_execution}). In parallel with the data deployment, Yorc orchestrates the deployment of the main workflow components in the computing infrastructures and manages their lifecycle (configuring, starting services) as described in the TOSCA part in the workflow description (Step 3a in Figure~\ref{fig:workflow_deployment}). 

This deployment is managed by means of containers that offer the simplest way to distribute software. However, the creation and deployment of the container images will differ depending on their functionality and target environment. For components and software deployed in the Cloud, we follow the traditional toolchain with images created from generic binary packages provided by the operating systems. In the case of HPC software, containers are built according to the target architecture of the HPC system in order to achieve the performance offered by these systems. In these cases, we propose a combination of the container image build procedures with HPC build systems like Spack~\cite{spack} or easybuild~\cite{easybuild}. These systems are used to manage the installation of software for HPC environments facilitating the installation of the software different compiler tool chains and architectures. Introducing them in the container images build procedures will bring the benefit of the container-based software distribution while fulfilling the requirements for getting good performance in HPC systems.

Regarding resource management, Yorc is in charge of orchestrating the resource provisioning by contacting the Cloud Manager (such as OpenStack) and deploying the containers as indicated in the TOSCA topology. In the case of HPC clusters, due to their connectivity and security constraints, the images are exported to files and transferred to the HPC storage, which will be deployed as containers at execution time using specialized HPC container engines such as Singularity~\cite{singularity}.

Once the workflow components and initial data have been deployed, Yorc submits the execution of the main workflow processes to the HPC infrastructure through UNICORE~\cite{unicore}, which is in charge of managing the federation of HPC compute and data resources in order to make them available to users in a secure way (Step 1 in Figure~\ref{fig:workflow_execution}). At the lowest level, the COMPSs runtime~\cite{compss_softwareX} will coordinate the invocations of the workflow components implemented with the PyCOMPSs task-based programming model (Step 2a in Figure~\ref{fig:workflow_execution}). As mentioned earlier, COMPSs supports several task types which can include HPC simulations, DA transformations, etc. The runtime dynamically generates a task-dependency graph by analysing the existing data dependencies between the invocations of tasks defined in the Python code. The task-graph encodes the existing parallelism of the workflow, which can be used to schedule the execution in the resources already deployed by Yorc. 
Based on this scheduling, the COMPSs runtime can interact with the different HPC, DA, and ML runtimes to coordinate resource usage, deciding which parts can run in parallel 
to deliver high overall performance.

However, the performance of the HPC simulations, DA algorithms, and/or training or inference DL processes is related to several input parameters that specify data partitionings and degrees of parallelism, such as the number of MPI processes, the data chunk size, and/or training batch. This is usually provided by the user and their optimal value is decided after a trial-and-error process. This user decision can be improved by an AI-assisted system where the optimal configuration (chunk size or number of processes) of the different workflow parts is inferred at runtime based on historical information. Each time a workflow is executed, the runtime systems (Yorc and COMPSs) store profiling data about the duration and resource usage of the different workflow tasks. This information, together with metadata for the dataset and the used execution configuration, can be used to train an ML model which relates the configuration to the used resources and the execution time. Then, every time a new execution of a workflow is submitted, the runtime systems can use this model to infer the best workflow configuration for the given the input dataset and the available resources.

Apart from the dynamic task graph generation, the COMPSs runtime offers several other features that enable different types of dynamicity in the workflow.
For example, it is able to react to task-failures and exceptions in order to adapt the workflow behaviour accordingly ~\cite{ejarque2020managing}. 
Moreover, it is also able to combine distinct types of data patterns in the same workflow. For instance, it supports data streams for communicating multiple tasks, or between tasks and the main code. It can be used to create workflows, whose computation is adapted to dynamic data sources, or to track the results of long-lasting computations before finishing the execution, allowing faster reactions by canceling unnecessary computations or spawning new ones with more relevant parameters~\cite{compss_streams}.

These functionalities, together with similar features provided by Yorc at a higher level, enables the possibility of supporting workflows with a very dynamic behaviour as described in Challenge 3.

Finally, with respect to the integration of the data management and computation, the eFlows4HPC stack provides two solutions for persistent storage: Hecuba (based on key-value databases) and dataClay (object-oriented distributed storage) \cite{dataClay}. These solutions can be leveraged by PyCOMPSs applications to store application objects as persistent objects in new memory devices such as NVRAM or SSDs, enabling the keeping of data after the execution of the application. This changes the paradigm of persistent storage in HPC, dominated by the file system, to more flexible approaches. By using persisted objects, application patterns such as producer-consumer, in-situ visualisation or analytics, can be easily implemented. 

Both solutions, Hecuba and dataClay, implement a common API that allows programmers to manipulate all the data as regular Python objects, regardless of whether they are persistent (stored in disk, NVRAM or similar) or volatile (stored in memory). They allow decoupling the data view of the user and the data organization in the storage system, which is defined according to the underlying storage system and transparently to the programmer. In addition, they can execute integrated with PyCOMPSs to enhance data locality and optimize the mechanism of passing parameters to tasks.
Hence, the proposed data abstraction layer implemented by Hecuba and dataClay addresses Challenge 4.


\subsection{Architectural Optimizations within eFlows4HPC}

An internal and important aspect within eFlows4HPC is the actual performance achieved for the execution of the workflows. Therefore, the project also puts the focus on the identification and optimization of the time-consuming kernels (understood as independent pieces of code with a well-defined functionality). The optimization process takes into account not only raw performance but also performance-per-Watt. Indeed, new energy-efficient heterogeneous architectures currently being deployed in HPC and DA ecosystems will be targeted for the workflow applications. The set of hardware solutions inspected in the project ranges from pre-exascale systems, such as MareNostrum 5, to high-end FPGA devices.

As an example, in the field of workflows using AI-specific components, the project focuses on developing specific kernel optimizations for heterogeneous architectures. One example is the optimization of convolution operations for DL training and inference. In this direction, a dataflow-oriented programming environment, such as that provided in HLS by Xilinx, enables the design of a pipeline-oriented kernel where throughput is maximized, latency is minimized and, in principle, higher energy efficiency is achieved.

A similar approach is followed in the direction of simulation-oriented HPC applications: specific kernels can be identified and optimized for new emerging technologies, such as RISC-V processors. In that aspect, the European Processor Initiative (EPI)~\cite{web:epi} is taken into account with the deployment of RISC-V architectures, emulated on the Marenostrum Exascale Emulation Platform (MEEP)~\cite{meep}.

Finally, GPU-based architectures are also considered for performance improvement of specific AI-related optimizations, mostly for distributed training as required by the project workflows.
\section{Discussion}
\label{sec:discusion}

One of they key aspects of our proposal is the concept of HPCWaaS. It applies the Function-as-a-Service (FaaS) model to the execution of complex workflows in hybrid HPC-Cloud systems. FaaS offers the functionality of exposing functions in the Cloud without dealing with the deployment and execution details in the Cloud. In this model, a function is registered to the FaaS platform in the Cloud and other users can invoke it from anywhere using a REST API. This deployment-execution model is the concept that we port to the HPC environment. However, there are important differences, such as the  computation granularity and the complexity in the workflow development and deployment. Current commercial or open-source FaaS approaches, which enable deployment on private data centres, are focused on fine-grain computations (duration is restricted to a few minutes) and do not support using multiple nodes to host a function execution. Moreover, the HPC environments are restricted environments, and system administrators are reluctant to install this type of services.

In order to tackle all these constraints, we propose to build HPCWaaS on top of the proposed software stack designed to operate with HPC environments instead of trying to adapt one of the existing Cloud FaaS tools.

A similar concept is the Workflow-as-a-Service (WaaS or WfaaS). It was initially introduced in~\cite{waas2009} and several proposals have been described in ~\cite{compss_servicess,WaaS2014,waas2015,2016waas,waas2018}. These proposals focus on migrating traditional scientific workflows as services in the Cloud. This migration has been performed in two ways: one dedicated to integrating the Workflow Management Systems (WMS) with Infrastructure-as-a-Service (IaaS) offerings where computations in the workflows are scheduled in VMs; and an alternative one focused on exposing the WMS features as a service, where users can submit the workflows and the system schedules its tasks in the available VMs.  In our approach, we deal with complex workflows that combine the invocation of several computational tasks. These computations require the integration of a variety of frameworks deployed and configured in hybrid distributed environments which are difficult to express in traditional workflow languages such as CWL or WDL. This is the reason why we propose a new way to describe workflows at different levels, where TOSCA is located in the upper level to specify the workflows deployment topology.
An alternative to indicate the workflow deployment with Kubernetes is proposed in the EVOLVE project\cite{web:evolve}. Despite this solution seeming to cover the same deployment requirements as our proposal, this container environment is rarely deployed in HPC clusters because it collides with the traditional HPC resource managers and queue systems. 

TOSCA and one of its orchestrators (Yorc) are also used in projects such as LEXIS\cite{lexis2020} and ACROSS\cite{web:accross} for similar functionalities. However, we propose a multi-level orchestration approach combining TOSCA with a task-based programming model (PyCOMPSs) and the Data Logistic Pipelines that allow developers to describe all the operations involved in a workflow. As explained in Section~\ref{sec:development},  TOSCA is used for high-level deployment orchestration, the Data Logistic Pipelines is used to describe data collection and integration, and PyCOMPSs for the lower-level dynamic workflow execution integrating in-situ AI/ML steps. This combined description enables the eFlows4HPC components to perform a reusable and fully automated workflow deployment and execution from scratch. 

Regarding dynamicity in scientific workflows, this aspect has been previously addressed with the concept of \textit{workflow steering}~\cite{wf_steering}. It consists of providing users with a kind of interactivity to perform fine-tune changes in the workflow during its execution. These changes are done by steering actions that may significantly improve the performance of the system reaching the same or better resource faster or using less resources. In our case, the PyCOMPSs programming model offers to developers the possibility to program these steering actions as part of the workflow and  automatically apply them during its execution. As explained in Section~\ref{sec:execution}, PyCOMPSs provides mechanisms for indicating failure reaction policies that inform the runtime what to do with the rest of the workflow when a task fails. It also provides exception management in parallel distributed workflows, which automatically cancels tasks and spawns new ones when an exceptional event occurs during the execution. Finally, PyCOMPSs allows streaming communication between different parts of a workflow that can be used to evaluate intermediate results and enables the implementation of in-situ optimization algorithms which can be combined with AI/ML techniques.

Regarding storage, the use of key-value and object store technologies under an object-oriented Python interface facilitate programmability of workflows, also avoiding serializations and deserializations to files in order to share data between different steps. Additionally, adding this data abstraction layer provides the ability to transparently optimize data access thanks to new storage devices, such as persistent memories. To add to a workflow already existing applications that use specific libraries or interfaces to access structured data (for example, HDF5 or NetCDF), our proposal is to integrate our storage technologies with these interfaces. With this approach, these applications could benefit from our software stack transparently to the programmer. However, to support applications with an access pattern that cannot benefit from our storage technologies, we also allow the utilization of traditional POSIX file systems, so the utilization of both types of storage systems can be combined in the same workflow.

\section{Conclusions}
\label{sec:conclusion}

In recent years HPC, along with DA and AI, has evolved providing the user community with powerful tools to tackle their application problems. However, the lack of programming environments for the development of workflows that include all three aspects is limiting their convergence. 

We have identified four main challenges 
that need to be overcome to achieve this convergence. First, the need for tools that foster openness, transparency, reusability, and reproducibility of the workflows and their results. Such tools are available in cloud environments but cannot be directly used in HPC systems. Therefore, new tools should be built, or adapted from existing ones, to offer these functionalities to HPC users, while complying with the constraints of HPC policies. Second, the development of complex workflows should be made easier while keeping their capabilities and performance. New methodologies are required to support the development of these workflows and simultaneously bridge the gap with the current and future heterogeneous infrastructures. 
Third, workflow managers must support dynamicity beyond static pipelines and simple static graphs. The engines should accommodate dynamic shifts in the requested computation according to changes in the input data, computation with urgency demands, and dynamic changes in the workflow executions due to eager analysis of results, exceptions or software faults. In addition, the engine should be able to leverage elastic resource management to deal with changes in the instant workload of the workflow. 
The fourth challenge comes from the data aspects and their integration with the computation. Similarly to the programming model, HPC and DA have relied on different solutions to store the data. Solutions that integrate the alternative data practices and offer an abstraction layer are necessary. Indeed, with the appearance of new storage devices, the solutions should leverage and aggregate them in this single data layer. 

Taking into account these challenges, we have proposed an architecture for a workflows software stack that offers tools to simplify the development, deployment, and execution of the type of complex workflows that we have described. In addition to the software stack, the HPC Workflows-as-a-Service (HPCWaaS) paradigm has been proposed as a mechanism to enable reuse, easy deployment, execution and reproduction of the workflow. The paradigm has been thought as a mechanism to lower the barrier toward the adoption of HPC systems and widen the access to a larger community of users. These ideas are under development in the EuroHPC eFlows4HPC project.

    \section*{Acknowledgements}
This work has received funding from the European High-Performance Computing Joint Undertaking (JU) under grant agreement No 955558. The JU receives support from the European Union’s Horizon 2020 research and innovation programme and Spain, Germany, France, Italy, Poland, Switzerland and Norway. In Spain, it has received complementary funding from MCIN/AEI/10.13039/501100011033, Spain and the European Union NextGenerationEU/PRTR (contracts PCI2021-121957, PCI2021-121931, PCI2021-121944, and PCI2021-121927). In Germany, it has received complementary funding from the German Federal Ministry of Education and Research (contracts 16HPC016K, 6GPC016K, 16HPC017 and 16HPC018). In France, it has received financial support from Caisse des dépôts et consignations (CDC) under the action PIA ADEIP (project Calculateurs). In Italy, it has been preliminary approved for complimentary funding by Ministero dello Sviluppo Economico (MiSE) (ref. project prop. 2659). In Norway, it has received complementary funding from the Norwegian Research Council, Norway under project number 323825. In Switzerland, it has been preliminary approved for complimentary funding by the State Secretariat for Education, Research, and Innovation (SERI), Norway. In Poland, it is partially supported by the National Centre for Research and Development under decision DWM/EuroHPCJU/4/2021. The authors also acknowledge financial support by MCIN/AEI /10.13039/501100011033 through the “Severo Ochoa Programme for Centres of Excellence in R\&D” (CEX2018-000797-S), the Spanish Government (contract PID2019-107255 GB) and by Generalitat de Catalunya (contract 2017-SGR-01414). Anna Queralt is a Serra Húnter Fellow.

    \bibliographystyle{elsarticle-num}
    \bibliography{eflows4hpc}

\end{document}